\documentclass[graybox]{svmult}

\usepackage{helvet}         
\usepackage{courier}        
\usepackage{type1cm}        
\usepackage{makeidx}         
\usepackage{graphicx}        
\usepackage{multicol}        
\usepackage[bottom]{footmisc}

\makeindex             

\usepackage{amsmath, amssymb, amsfonts}
\usepackage{amsxtra, latexsym, amscd}
\usepackage{color}

\newcommand{\cA}{{\mathcal A}}

\newcommand{\cB}{{\mathcal B}}
\newcommand{\cD}{{\mathcal D}}

\newcommand{\cP}{{\mathcal P}}
\newcommand{\cR}{{\mathcal R}}
\newcommand{\sAA}{{\mathbb A}}
\newcommand{\sBB}{{\mathbb B}}
\newcommand{\sCC}{{\mathbb C}}
\newcommand{\nn}{{\mathbb N}}
\newcommand{\ff}{{\mathbb F}}

\newcommand{\bldg}{{\mbox{\bf g}}}

\newcommand{\bldx}{{\mbox{\bf x}}}
\newcommand{\bldy}{{\mbox{\bf y}}}

\newcommand{\code}{\mathcal{C}}

\newcommand{\blde}{{\mbox{\bf e}}}

\newcommand{\bldG}{{\mbox{\bf G}}}
\newcommand{\bldI}{{\mbox{\bf I}}}

\newcommand{\bldA}{{\mbox{\bf A}}}

\begin{document}
\title*{Batch and PIR Codes and Their Connections to Locally Repairable Codes}

\author{Vitaly Skachek}
\institute{The author is with the Institute of Computer Science, University of Tartu, Tartu 50409, Estonia,  
\email{vitaly.skachek@ut.ee}}

\maketitle

\abstract{
Two related families of codes are studied: batch codes and codes for private information retrieval. 
These two families can be viewed as natural generalizations of locally repairable codes, which were extensively studied in the context of coding for fault tolerance in distributed data storage systems. Bounds on the parameters of the codes, as well as 
basic constructions, are presented. Connections between different code families are discussed. 
}

\section{Introduction}

In this survey, we discuss two related families of codes: batch codes and codes for private information retrieval (PIR codes). 
These two families can be viewed as natural generalizations of locally repairable codes, which were extensively studied in the context of coding for fault tolerance in distributed data storage systems. 

Batch codes were first presented in~\cite{Ishai}, where it was suggested to use them for load balancing in the multi-server distributed data storage systems. It was also suggested in~\cite{Ishai} to use these codes in private information retrieval. 
A number of constructions of batch codes were presented therein. 
Later, the authors of~\cite{Wang2013} proposed to use so-called ``switch codes'' for facilitating the routing of data in the network switches. It turns out, however, that switch codes are a special case of batch codes. 

Coding schemes for PIR were studied in~\cite{Fazeli}. The authors showed that a family of codes, 
which is a relaxed version of batch codes, can be employed in classical linear PIR schemes in order to reduce the redundant information stored in a distributed server system. This relaxed version of the batch codes is termed PIR codes.  

In these schemes, typically a distributed data storage system is considered. The coded words are written 
across the block of disks (servers), where each disk stores a single symbol (or a group of symbols). 
The reading of data is done by accessing a small number of disks. Mathematically, this can 
be equivalently represented by the assumption that each information symbol depends on a small number of other symbols. 
However, the type of requested queries varies in different code models. Thus, in PIR codes several copies of the 
same information symbols are requested, while in batch codes combinations of different symbols are also possible. 

In this survey, we mathematically define the corresponding families of codes, and study their properties. 
We derive bounds on the parameters of such codes, and show some basic constructions. 
We also show relations between different families of codes. 
In Section~\ref{sec:LRC-codes}, we introduce various models of locally repairable codes. 
In Section~\ref{sec:batch-codes}, we define batch codes. 
In Section~\ref{sec:linear}, we discuss properties of linear batch codes. 
In Section~\ref{sec:PIR-codes}, we introduce codes for private information retrieval. 
In Section~\ref{sec:connections}, we study connections  between locally repairable and batch/PIR codes.
In Sections~\ref{sec:bounds-unrestricted} and~\ref{sec:bounds-restricted}, 
we present bounds on the parameters of various families of codes. 
In Section~\ref{sec:open-questions}, we pose some open questions. 
For the sake of completeness, we introduce all necessary notations and definitions, which are used in the sequel. 

\section{General Settings}

We denote by $\nn$ the set of nonnegative integer numbers. For $n \in \nn$, define $[n] \triangleq \{1, 2, \cdots, n\}$. 
Let $\blde_i$ be the row
vector having one at position $i$ and zeros elsewhere (the length of vectors will be clear from the context). 

Let $\Sigma$ be a finite alphabet. Let $\bldx = (x_1, x_2, \cdots, x_k) \in \Sigma^k$ be an information vector. 
The code is a set of coded vectors 
\[
\left\{ \bldy = (y_1, y_2, \cdots, y_n) = \code(\bldx) \; : \; \bldx \in \Sigma^k \right\} \subseteq \Sigma^n \; , 
\]
where $\code : \Sigma^k \rightarrow \Sigma^n$ is a bijection, for some $n \in \nn$. 
By slightly abusing the notation, sometimes we denote the above set by $\code$. 

Let $\ff = \ff_q$ be a finite field with $q$ elements, where $q$ is a prime power. 
If $\code : \ff^k \rightarrow \ff^n$ is a linear mapping, then $\code$ is a linear $[n,k,d]$ code over $\ff$.
Here, $d$ is the minimum Hamming distance of $\code$. In that case, the encoding can be viewed as a multiplication 
by a $k \times n$ generator matrix $\bldG$ over $\ff$ of an information vector $\bldx$, 
\begin{equation}
\bldy = \bldx \cdot \bldG \; . 
\label{def:linear}
\end{equation}

The rate of the code is defined as $\cR \triangleq k/n$. 

\section{Codes with Locality and Availability}
\label{sec:LRC-codes}

Codes with locality were proposed for use in the distributed data storage systems~\cite{dimakis-survey}. 
In such systems, the data is stored in many disks (servers), and these servers may fail from time to time. 
It is possible to use erasure-correcting codes, where parts of a codeword are stored in different servers.
A failure of one server can be viewed as an erasure of a symbol or of a group of symbols. 
In order to repair the erased symbols, there is a need to bring a few other symbols from other servers. 
In general, it would be beneficial to minimize the traffic in the network. 
We continue by recalling the definition of \emph{locally repairable codes} (LRCs).  

\begin{definition}
The code $\code$ has locality $r \ge 1$, if for any $\bldy \in \code$,
any symbol in $\bldy$ can be recovered by using \emph{at most $r$} other symbols of $\bldy$.
\end{definition}

Codes with low locality were extensively discussed, for example, in~\cite{dimakis-survey}. 
A somewhat similar family of codes, known as \emph{one-step majority-logic decodable codes}, was investigated in the classical literature~\cite{Massey}. 
Recently, the bounds on the parameters of LRCs were derived in~\cite{Yekhanin2012}. It was shown therein that 
the parameters of a linear $[n,k,d]$ code with locality $r$ over $\ff$ satisfy:
\begin{equation}
n \ge k + d + \Bigg\lceil \frac{k}{r} \Bigg\rceil - 2 \; . 
\label{eq:bound-gopalan}
\end{equation}
This bound can be viewed as a refinement of the classical Singleton bound, where $\lceil \frac{k}{r} \rceil - 1$ is an additive penalty for locality of the code, when compared to the classical Singleton bound. The proof is done by iterative expurgating
of the code, and by taking into account that there are dependencies between sets of $r + 1$ symbols, which consist of an arbitrary symbol in $\bldy$ and its recovery set. The bound in~(\ref{eq:bound-gopalan}) is tight. In fact, several known constructions attain it with equality (see, for example, ~\cite{Yekhanin2012, Silberstein-construction, Tamo-construction}).  
\medskip

Assume that the linear code $\code$ is systematic, i.e. the matrix $\bldG$ contains a $k \times k$ identity submatrix $\bldI$. 
Then, the symbols of $\bldy$ corresponding to $\bldI$ are called \emph{information symbols}.  
It is possible to require recoverability of information symbols only (from sets of size at most $r$). In that case, the code is said to have \emph{locality of information symbols}. Otherwise, if \emph{all} symbols of $\bldy$ are recoverable from small sets, the code has \emph{locality of all symbols}. 

\medskip
The above model was extended to codes with \emph{locality and availability} in~\cite{RPDV2014}. 
\begin{definition}
The code $\code$ has locality $r \ge 1$ and availability $\delta \ge 1$, if for any $\bldy \in \code$, 
any symbol in $\bldy$ can be reconstructed by using any of $\delta$ disjoint sets of symbols, each set is of size at most $r$.
\end{definition}
In~\cite{wang-zhang}, the authors consider linear codes with locality $r$ and availability $\delta$ of all symbols. They derive the following bound on the parameters of the code: 
\begin{equation}
n \ge k + d + \Bigg\lceil \frac{\delta(k-1)+1}{\delta(r-1)+1} \Bigg\rceil - 2 \; . 
\label{eq:bound-wz}
\end{equation}
In~\cite{RPDV2014}, systematic codes (linear or non-linear) are considered, with locality $r$ and availability $\delta$ of information symbols. The authors show the bound analogues to~(\ref{eq:bound-wz}) for that case. 
In particular, we observe that when availability $\delta = 1$, i.e. there is only one recovery set for each symbol, then~(\ref{eq:bound-wz}) coincides with~(\ref{eq:bound-gopalan}). The proof technique in both cases is based on the idea similar to that of~\cite{Yekhanin2012}. 
\medskip 

Another related model is considered in~\cite{RMV2014}. In that model, several \emph{different} symbols are recovered from a small set of recovery symbols. By building on the ideas in the previous works, the authors derive a variation of the bound~(\ref{eq:bound-wz}) for the model under consideration. Other related works, for example, 
include~\cite{Cadambe, Paudyal, Prakash, Tamo, Hollanti15, Hollanti16} and the references therein.

\section{Batch Codes}
\label{sec:batch-codes}

Batch codes were first presented in the cryptographic community in~\cite{Ishai}. In that work, the authors proposed to use batch codes for load balancing in the distributed systems, as well as for private information retrieval. The authors of~\cite{Ishai}
have also presented a few constructions of various families of batch codes. Those constructions were based on recursive application of simple batch codes (so-called ``sub-cube codes''), on classical Reed-Muller codes, on locally decodable codes, and others. 
 
The following definition is based on~\cite{Ishai}. 
\begin{definition}
Let $\Sigma$ be a finite alphabet. We say that $\code$ is an $(k, n, t, M, \tau)_\Sigma$ \emph{batch code} over a finite alphabet $\Sigma$ if it encodes any string $\bldx = (x_1, x_2, \cdots, x_k) \in \Sigma^k$ into $M$ strings (buckets) of total length $n$ 
over $\Sigma$, namely $\bldy_1, \bldy_2, \cdots, \bldy_M$,  such that for each $t$-tuple (batch) of (not neccessarily distinct)
indices $i_1, i_2, \cdots, i_t \in [k]$, the symbols
$x_{i_1}, x_{i_2}, \cdots, x_{i_t}$ can be retrieved by reading at most $\tau$ symbols from each bucket. 
\label{ref:batch-def}
\end{definition}

More formally, by following the presentation in~\cite{Vardy}, we can state an equivalent definition.
\begin{definition}
An $(k, n, t, M, \tau)_\Sigma$ batch code $\code$ over a finite alphabet $\Sigma$ is defined by
an encoding mapping $\code \; : \; \Sigma^k \rightarrow (\Sigma^*)^M$ (there are $M$ buckets in total), and a decoding mapping $\cD \; : \; \Sigma^n \times [k]^t\rightarrow \Sigma^t$, such that  
\begin{enumerate}
\item
The total length of all buckets is $n$; 
\item
For any $\bldx \in \Sigma^k$ and 
\begin{equation}
i_1, i_2, \cdots, i_t \in [k] \; , 
\label{eq:querry}
\end{equation}
\[
\cD\left(\code(\bldx), i_1, i_2, \cdots, i_t\right) = (x_{i_1}, x_{i_2}, \cdots, x_{i_t}) \; ,  
\]
and $\cD$ depends only on $\tau$ symbols in each bucket in $\code(\bldx)$. 
\end{enumerate}
\end{definition}

In particular, an interesting case for consideration is when $\tau = 1$, namely only at most one symbol is read from each bucket.  
If the requested information symbols 
$(x_{i_1}, x_{i_2}, \cdots, x_{i_t})$ can be reconstructed from the data read by $t$ different users independently
(i.e., the symbol  $x_{i_\ell}$ is reconstructed by the user $\ell$, $\ell = 1, 2, \cdots, t$, respectively), 
and the sets of the symbols read by these $t$ users are all disjoint, 
such a model is called a \emph{multiset} batch code. 

In the sequel, we only consider multiset batch codes, 
and therefore we usually omit the word ``multiset'' for convenience. 
\medskip 

An important special case of batch codes is defined as follows. 
\begin{definition}[\cite{Ishai}]
A \emph{primitive} batch code is a batch code, where each bucket contains exactly one symbol. In particular, $n = M$. 
\end{definition}
\medskip

Following the work~\cite{Ishai}, a number of subsequent papers have studied \emph{combinatorial} batch codes. In combinatorial batch codes, a number of replicas of the information symbols are stored in different positions in the codeword. Usually, the symbols are associated with servers according to some optimal or sub-optimal combinatorial objects, such as block designs. 
Combinatorial batch codes were studied, for example, in~\cite{Bhattacharya,Brualdi,Bujtas,Gal,Stinson}.

\section{Linear Batch Codes}
\label{sec:linear}

In what follows, we consider a special case of primitive multiset batch codes with $n = M$ and $\tau=1$. 
Under these conditions, each symbol can be viewed as a separate bucket, and only one reading per bucket is allowed. 

We assume that the information and the coded symbols are taken from the finite field $\ff = \ff_q$, where $q$ is a prime power. 
Additionally, we assume that the encoding mapping $\code : \ff^k \rightarrow \ff^n$ is linear over $\ff$, and therefore the code $\code$ is a linear $[n,k]$ code over $\ff$. In that case, $\code$ falls under 
the linear coding framework defined in~(\ref{def:linear}). We also refer to the parameter $t$ as the size of a query of the code.  
The batch code with the parameters $n$, $k$ and $t$ over $\ff_q$ is denoted as $[n,k,t]_q$-batch code 
(or simply $[n,k,t]$-batch code) in the sequel. 

This framework was first considered in~\cite{Lipmaa}, and the similarities with locally repairable codes were mentioned. 
The main difference between these two families, however, is that the supported query types are different. In batch codes we are interested in reconstruction of the information symbols in $\bldx$, while in locally repairable codes 
the coded symbols in $\bldy$ are to be recovered. 

The following simple result was established in~\cite[Theorem 1]{Lipmaa}.

\begin{theorem}
Let $\code$ be an $[n, k, t]_q$ batch code. It is possible to retrieve $x_{i_1}, x_{i_2}, \cdots, x_{i_t}$ 
by $t$ different users in the primitive multiset batch code model
(where the symbol  $x_{i_\ell}$ is retrieved by the user $\ell$, $\ell = 1, 2, \cdots, t$, respectively)
if and only if there exist $t$ non-intersecting sets $T_1, T_2, \cdots, T_t$ of indices of columns 
in the generator matrix {$\bldG$}, 
and for each $T_\ell$, $1 \le \ell \le t$, there exists a linear combination of columns of {$\bldG$} indexed by that set, 
which equals to the column vector $\blde_{i_\ell}^T$, for all $\ell \in [t]$.
\label{thrm:lipmaa-1}
\end{theorem}

The reader can find the proof of this theorem in~\cite{Lipmaa}. Next, we show examples that further illustrate this concept.  

\begin{example}[\cite{Ishai}]
Consider the following binary $2 \times 3$ generator matrix of a batch code $\code$ given as 
\[
\bldG = \left( 
\begin{array}{ccc}
1 & 0 & 1  \\
0 & 1 & 1  \\
\end{array}
\right) \; . 
\]
The corresponding code is a sub-cube code constructed in~\cite[Section 3.2]{Ishai}. By using this code, 
the information symbols $(x_1, x_2)$ are encoded into $(y_1, y_2, y_3) = (x_1, x_2, x_1 + x_2)$. 

Assume that the query contains two different symbols $(x_1, x_2)$. 
Then, we can retrieve these symbols direclty by using the following equations: 
\[
\left\{ 
\begin{array}{ccl}
x_1 & = & y_1 \\
x_2 & = & y_2 
\end{array} \right. \; . 
\]
Alternatively, assume that the query contains two copies of the same symbol, for example $(x_1, x_1)$. 
Then, we can retrieve these symbols by using the following equations: 
\[
\left\{ 
\begin{array}{ccl}
x_1 & = & y_1 \\
x_1 & = & y_2 + y_3
\end{array} \right. \; . 
\] 
Similarly, $(x_2, x_2)$ can be retrieved. We conclude that $\code$ is a $[3, 2, 2]_2$ batch code. 
\end{example}

\begin{example}[\cite{Lipmaa}]
Pick the following binary $4 \times 9$ generator matrix of a batch code $\code$ given as 
\[
\bldG = \left( 
\begin{array}{ccccccccc}
1 & 0 & 1 & 0 & 0 & 0 & 1 & 0 & 1 \\
0 & 1 & 1 & 0 & 0 & 0 & 0 & 1 & 1 \\
0 & 0 & 0 & 1 & 0 & 1 & 1 & 0 & 1 \\  
0 & 0 & 0 & 0 & 1 & 1 & 0 & 1 & 1 
\end{array}
\right) \; . 
\]
The corresponding code is a second-order sub-cube code constructed as in~\cite[Section 3.2]{Ishai}.

Assume that the query contains the information symbols $(x_1, x_1, x_2, x_2)$. 
Then, we can retrieve these symbols using the following equations: 
\[
\left\{ 
\begin{array}{ccl}
x_1 & = & y_1 \\
x_1 & = & y_2 + y_3 \\
x_2 & = & y_5 + y_8 \\
x_2 & = & y_4 + y_6 + y_7 + y_9 
\end{array} \right. \; . 
\]
It can be verified in a similar manner that any $4$-tuple $(x_{i_1}, x_{i_2}, x_{i_3}, x_{i_4})$, where $i_1, i_2, i_3, i_4 \in [4]$, can be retrieved by using the symbols of $\bldy$, by using each symbol at most once. 
We conclude that $\code$ is a $[9, 4, 4]_2$ batch code. 
\end{example}

\begin{example}[\cite{WKC2015}]
\label{ex:simplex}
Pick the following binary $3 \times 7$ generator matrix of a batch code $\code$ given as 
\[
\bldG = \left( 
\begin{array}{ccccccccc}
1 & 0 & 0 & 1 & 1 & 0 & 1 \\
0 & 1 & 0 & 1 & 0 & 1 & 1 \\
0 & 0 & 1 & 0 & 1 & 1 & 1 
\end{array}
\right) \; . 
\]
The corresponding code is a binary $[7,3,4]$ classical error-correcting simplex code.

Assume that the query contains the information symbols $(x_1, x_1, x_2, x_2)$. 
Then, we can retrieve these symbols using the following equations: 
\[
\left\{ 
\begin{array}{ccl}
x_1 & = & y_1 \\
x_1 & = & y_2 + y_4 \\
x_2 & = & y_3 + y_6 \\
x_2 & = & y_5 + y_7 
\end{array} \right. \; . 
\]
It can be verified in a similar manner that any $4$-tuple $(x_{i_1}, x_{i_2}, x_{i_3}, x_{i_4})$, where $i_1, i_2, i_3, i_4 \in [4]$, can be retrieved by using the symbols of $\bldy$, by using each symbol at most once. 
We conclude that $\code$ is a $[7, 3, 4]_2$-batch code. Moreover, it was shown in~\cite{WKC2015} that all queries can be 
satisfied when each user reads at most $r=2$ symbols from $\bldy$.  
\end{example}

A family of codes, related to batch codes, and corresponding to the case $t=k$, termed \emph{switch codes}, 
was studied in \cite{Chee, Wang2013, WKC2015}. It was suggested in~\cite{Wang2013} to use such codes for efficient 
routing of data in the network switches. 
\medskip 

The following property of batch codes was observed in~\cite{Lipmaa} for binary linear codes, and later generalized to nonbinary
(and also to non-linear) codes in~\cite{Zumbragel}. 

\begin{theorem}
Let $\code$ be an $[n, k, t]_q$-batch code. Then, the minimum Hamming distance of $\code$ is at least $t$. 
\label{thrm:dist-batch}
\end{theorem}

{\bf Proof.}
Let $\bldy_1 = \code(\bldx_1)$ and $\bldy_2 = \code(\bldx_2)$ be two codewords of $\code$, and $\bldx_1 \neq \bldx_2$. 
Then, $\bldx_1$ and $\bldx_2$ differ in at least one symbol, i.e. $(\bldx_1)_\ell \neq (\bldx_2)_\ell$, for some $1 \le \ell \le k$. Consider the query
$(\underbrace{x_\ell, x_\ell, \cdots, x_\ell}_{t})$. The $i$-th copy of $x_\ell$ is recovered from the set of symbols indexed by the set $T_i$, $1 \le i \le t$. Since $\bldx_1$ and $\bldx_2$ differ in the $\ell$-th symbol, the codewords $\bldy_1$ and $\bldy_2$ should differ in at least one symbol in each $T_i$, $1 \le i \le t$. The sets $T_i$ are all disjoint, and therefore 
$\bldy_1$ and $\bldy_2$ differ in at least $t$ symbols. 
\qed

It follows that any $[n, k, t]_q$-batch code is in particular a classical $[n, k, \ge t]_q$ error-correcting code, 
and a variety of classical bounds, such as Singleton bound, Hamming bound, Plotkin bound, Griesmer bound, Johnson bound, Elias-Bassalygo bound, are all applicable to batch codes (when the minimum distance $d$ 
is replaced by the query size $t$). 

\medskip 

\begin{example}
Let $m>1$ be an integer. A binary simplex $[2^m-1, m, 2^{m-1}]$ code $\code$ is defined 
by its generator matrix 
\[
\bldG = \left( \bldg_1 \, | \, \bldg_2 \, | \, \cdots \, | \,\bldg_{2^m-1} \right) \; , 
\]
where $\bldg_i$ are all possible different binary nonzero column vectors of length $m$, $i = 1, 2, \cdots, 2^m-1$
~\cite[Problem 2.18]{Roth}. 

For a classical error-correcting $[n,k,d]$ code over $\ff_q$, the Plotkin bound is defined as 
follows~\cite[Theorem 2.2.29]{Pellikaan}: 
\[
\mbox{if $qd > (q-1)n$, \; then } \quad q^k \le \Bigg\lfloor \frac{qd}{qd - (q-1)n} \Bigg\rfloor \; . 
\]
 
It is straightforward to see that, as an error-correcting code, the binary simplex code as above (with $q = 2$) 
attains the Plotkin bound with equality~\cite[Problem 2.18]{Roth} for all $m \ge 2$. 

As it was shown in~\cite{WKC2015}, the code $\code$ is a $[2^m-1, m, 2^{m-1}]_2$ batch code, with $t = 2^{m-1}$. 
Therefore, by Theorem~\ref{thrm:dist-batch}, it attains the corresponding Plotkin-based bound
\[
q^k \le \Bigg\lfloor \frac{qt}{qt - (q-1)n} \Bigg\rfloor \; 
\]
with equality, and therefore it is a Plotkin-optimal batch code.  
\end{example}

In~\cite{Zhang-Skachek}, a variation of batch codes \emph{with restricted size of reconstruction sets} is defined. 
These codes are batch codes as in Definition~\ref{ref:batch-def} with an additional property that every queried information symbol $x_i$ is reconstructed from \emph{at most $r\ge 1$} symbols of $\bldy$.  
This additional property can be viewed as analogous to locality of the LRCs.  
Small size of reconstruction sets allows for recovering the requested data symbol from a small number $r$ of servers, 
thus reducing the traffic and the load in the system. 
\smallskip

For example, the binary simplex code $\code$ in the previous example was shown in~\cite{WKC2015} to have the size of reconstruction 
sets of at most $r=2$. 
\medskip

\section{Codes for Private Information Retrieval}
\label{sec:PIR-codes}

The topic of private information retrieval (PIR) protocols has been a subject of a lot of research over the last two decades~\cite{Chor}. In the PIR scenario, the database is stored in a number of servers in a distributed manner. The user is interested in reading an item from the database without revealing to any server what item was read. In the classical approach, the data is replicated, and the replicas are stored in a number of different servers. The user accesses some of these servers, such that no server learns what data the user is interested in (it is assumed that the servers do not collude). 

A novel approach to PIR is based on coding, and it was studied, for example in \cite{augot, chan, Kopparty}. 
More specifically, assume that $\bldx = (x_1, x_2, \cdots, x_k)$ is an information vector, which is encoded into 
$\code(\bldx) = \bldy = (y_1, y_2, \cdots, y_n)$. The symbols of $\bldy$ are stored in different servers
in a distributed manner. 

In~\cite{chan}, the authors show that there is a fundamental trade-off between download communication complexity of the protocol (the number of symbols or bits downloaded by the user from the database) and the storage overhead (the number of redundant symbols or bits stored in the database). Later, the authors of~\cite{Fazeli} show that it is possible to emulate many
of the existing PIR protocols by using a code of length $n$  that approaches $(1 + \epsilon)k$ for vanishing $\epsilon$
(for sufficiently large $k$). This approach leads to PIR schemes with storage data rate arbitrarily close to $1$. 
The authors define a special class of codes, which allows for such efficient PIR protocols.

\begin{definition}(\cite{Fazeli})
An $k \times n$ binary matrix $\bldG$ has property $\cA_t$
if for all $i \in [k]$, there exist $t$ disjoint sets of columns of $\bldG$ that add up to $\blde_i$. 
A binary linear $[n,k]$ code $\code$ is called a \emph{$t$-server PIR code} (or, simply, PIR code) if there
exists a generator matrix $\bldG$ for $\code$ with property $\cA_t$.
\end{definition}

The batch codes turn out to be a special case of PIR codes, with a difference that PIR codes support only queries of type $(x_i, x_i, \cdots, x_i)$, $i \in [k]$, while batch codes support queries of a more general form
$(x_{i_1}, x_{i_2}, \cdots, x_{i_t})$, possibly for different indices $i_1, i_2, \cdots, i_t \in [k]$. 
It follows that batch codes can be used as PIR codes.  

Since for PIR codes (as well as for batch codes), the code rate $\cR$ approaches $1$~\cite{Ishai, Rao-Vardy} 
for large values of $k$,
it is more appropriate to talk about 
redundancy (as a function of $k$), 
rather than about the code rate. This is in contrast to PIR/batch codes with restricted 
size of reconstruction sets, where the asymptotic loss of code rate takes place. 
The redundancy of the codes will be defined and analyzed in the following sections. 

Constructions of PIR codes were presented very recently, for example, in~\cite{etzion-blackburn, Fazeli, Ge}.

\section{Connections Between Batch/PIR Codes and General LRCs}
\label{sec:connections}

As it was mentioned above, there are two types of LRCs considered in the literature: LRCs with locality of information symbols and LRCs with locality of all symbols. 

In order to preserve the information symbols in the coded word, a code with locality of information symbols has a systematic encoding matrix $\bldG = [\bldI | \bldA]$ for some matrix $\bldA$. Consider LRCs with locality $r$ and availability $\delta = t-1$ of information symbols. Then, each information symbol $y_i$, $1 \le i \le k$, in $\bldy$, can be recovered from $\delta$ disjoint sets $T_i$ of symbols, $|T_i| \le r$.    
Such a code can also be viewed as a PIR code that supports any query of $t$ copies of an information symbol with locality $r$
(including one copy of the information symbol in the systematic part).   
Generally, it does not follow that such a code is a batch code, since there is no guarantee that mixed queries of different information symbols are 
supported by disjoint reconstruction sets. 

On the other hand, a systematic batch or PIR code with restricted size of reconstruction sets allows to recover any query of $t$ information symbols with recovery sets of 
size $r$. 
Since in the systematic case, the information symbols are a part of a coded word $\bldy$, it follows that 
this code is an LRC with locality $r$ and availability $\delta = t-1$ of information symbols.

We obtain the following corollary (see also Theorem~21 in~\cite{Fazeli}). 

\begin{corollary}
A linear systematic code $\code$ is an LRC with locality $r$ and availability $\delta = t-1$ of information symbols if and only if $\code$ is a PIR code that supports queries of size $t$ with size of reconstruction sets at most $r$. 
\end{corollary}

It follows that the bounds derived for the parameters of the LRCs with \emph{locality and availability of information symbols}
can be applied also to systematic batch or PIR codes.\footnote{Please note that generally it does not follow here that the bounds for LRCs with \emph{locality of all symbols} are applicable to systematic batch or PIR codes.} 

On the other hand, for the non-systematic case, there is no simple known connection between linear batch codes with restricted size of reconstruction sets and LRCs with availability, as it is illustrated in the following examples. 

\begin{example}
Let $\bldG$ be a $k \times (3k)$ generator matrix of a linear binary code $\code$ defined as follows:
\[
   \bldG = \left( \begin{array}{ccccc|ccccc|ccccc}
    1 & 0 & \ldots & 0 & 1 & 1 & 0 & \ldots & 0 & 1 & 1 & 0 & \ldots & 0 & 1 \\
    0 & 1 & \ldots & 0 & 1 & 0 & 1 & \ldots & 0 & 1 & 0 & 1 & \ldots & 0 & 1 \\
    0 & 0 & \ddots & 0 & 1 & 0 & 0 & \ddots & 0 & 1 & 0 & 0 & \ddots & 0 & 1 \\
		0 & 0 & \ldots & 1 & 1 & 0 & 0 & \ldots & 1 & 1 & 0 & 0 & \ldots & 1 & 1 \\
		0 & 0 & \ldots & 0 & 1 & 0 & 0 & \ldots & 0 & 1 & 0 & 0 & \ldots & 0 & 1 \\
		\end{array} \right) \; . 
\]
Specifically, the binary information vector $\bldx = (x_1, x_2, \ldots, x_k)$ is encoded into the codeword 
$\bldy$, which consists of three copies of the same sub-vector, 
\begin{eqnarray*}
\bldy & = & (y_1, y_2, \ldots, y_{3k}) \\
& = & (x_1, x_2, \ldots, \sum_{i=1}^k x_i, \; \; x_1, x_2, \ldots, \sum_{i=1}^k x_i, \; \;
x_1, x_2, \ldots, \sum_{i=1}^k x_i ) \; .
\end{eqnarray*}

The code $\code$, when viewed as an LRC, has locality $r = 1$ and availability $\delta = 2$, since every symbol in $\bldy$ can be recovered from a single symbol in $\bldy$, and there are 2 different recovery sets. 

On the other hand, the code $\code$, when viewed as a batch or PIR code, must have a maximal size of reconstruction sets at least $k$, since it is impossible to recover a single information symbol $x_k$ from less than $k$ coded symbols in $\bldy$. 
\end{example}

The following example shows that batch or PIR code with small size of reconstruction sets 
is not necessarily LRC with all symbols locality,
even in the systematic case (the example can be also modified to a non-systematic case). 

\begin{example}
Take $\bldG$ to be a binary $2 \kappa \times (3 \kappa + 1)$ matrix, where $\kappa$ is some integer, as follows:
\[
   \bldG = \left( \begin{array}{ccc|ccc|c|ccc|c}
    1 & 1 & 0 & 0 & 0 & 0 & \ldots & 0 & 0 & 0 & 1 \\
    0 & 1 & 1 & 0 & 0 & 0 & \ldots & 0 & 0 & 0 & 1 \\
    \hline
		0 & 0 & 0 & 1 & 1 & 0 & \ldots & 0 & 0 & 0 & 1 \\
		0 & 0 & 0 & 0 & 1 & 1 & \ldots & 0 & 0 & 0 & 1 \\
    \hline
		\vdots & \vdots & \vdots & \vdots & \vdots & \vdots & \ddots & \vdots & \vdots & \vdots & \vdots \\
		\hline
		0 & 0 & 0 & 0 & 0 & 0 & \ldots & 1 & 1 & 0 & 1 \\
    0 & 0 & 0 & 0 & 0 & 0 & \ldots & 0 & 1 & 1 & 1 \\
		\end{array} \right) \; . 
\]
Here, $k = 2 \kappa$. This matrix $\bldG$ is a diagonal block matrix, where each block is a $2 \times 3$ generator matrix of a basic sub-cube code in~\cite{Ishai}, with an additional all-ones column. Let $\code$ be a binary linear code generated by this matrix.  
\smallskip

The code $\code$, when viewed as a batch code, supports any two queries of the form $(x_i, x_j)$ ($1 \le i \le k$, $1 \le j \le k$), with size of reconstruction sets at most 2. Therefore, $\code$ is a batch code with $r=2$, $t=2$. In particular, it is a PIR code with $r=2$ and $t=2$.  

On the other hand, the code $\code$, when viewed as an LRC, has locality of at least $\kappa$, since in order to recover $y_n = \sum_{i=1}^k x_k$, one needs to combine at least $\kappa$ other symbols of $\bldy$. 

As we see, this batch code with $r=2$ and $t=2$ has locality at least $k/2 = \kappa$, when used as an LRC with all symbols locality.   
\end{example}

\section{Bounds on the Parameters of PIR and Batch Codes with Unrestricted Size of Reconstruction Sets}
\label{sec:bounds-unrestricted}

In~\cite{Vardy}, the systematic batch codes with \emph{unrestricted} size of reconstruction sets are considered. 
The codes under consideration have restriction on the value of $t$ (typically, it is a small constant),
yet there is no restriction on $r$, so we can assume that $r=n$. 

Define the parameters $\cB(k,t)$ and $\cP(k,t)$ to be the shortest length $n$ of any linear systematic batch and PIR code, respectively, 
for given values of $k$ and $t$. Define the \emph{optimal redundancy} of systematic batch and PIR codes, respectively, as
\[
r_B(k,t) \triangleq \cB(k,t)- k \quad \mbox{ and } \quad r_P(k,t) \triangleq \cP(k,t)- k \; . 
\] 
It is known~\cite{Vardy} that for any fixed $t$,
\[
\lim_{k \rightarrow \infty} \frac{\cB(k,t)}{k} = 1 \; . 
\]

In a case of switch codes, $k=t$, it is shown in~\cite{Wang2013} that $\cB(k,k) = O\left(k^2 / \log(k) \right)$. 
A constructive proof showing that $r_P(k,t) = t \cdot \sqrt{k} (1 + o(1))$ was given in~\cite{Fazeli}. 
As for the lower bound on $r_P(k,t)$, it was recently shown in~\cite{Rao-Vardy} (see also~\cite{Wootters}) 
that for a fixed $t \ge 3$, $r_P(k,t) = \Omega(k)$, thus establishing an asymptotic behavior for the redundancy of the    
PIR codes.  

Since every batch code is also a PIR code, it follows that $\cB(k,t) \ge \cP(k,t)$, and $r_B(k,t) \ge r_P(k,t)$.  
The relations between $\cB(k,t)$ and $\cP(k,t)$ for specific choices of $t$ were extensively studied in~\cite{Vardy}. 
Thus, for example, it was shown that $\cB(k,t) = \cP(k,t)$ for $1 \le t \le 4$, while for $5 \le t  \le 7$, 
\[
 r_B(k,t) \le r_P(k,t) + 2 \lceil \log (k) \rceil \cdot r_P(k/2, t-2) \; . 
\]
It is quite straightforward to verify that $r_B(k,1) = 0$ and $r_B(k,2) = 1$ for any $k$.  
It was additionally shown in~\cite{Vardy} that 
\begin{eqnarray*}
 r_B(k,t) & = & O(\sqrt{k}) \mbox{ for } t = 3 , 4 \; , \\
 r_B(k,t) & = & O(\sqrt{k} \cdot \log(k) ) \mbox{ for } 5 \le t \le 7 \; . 
\end{eqnarray*} 

The following more general result was proven in~\cite{Vardy}. 

\begin{theorem} 
\label{thrm:redundancy}
For all values of $k$ and $t$, it holds
\[
r_B(k,t) \le r_P(k,t) + \Big\lfloor \frac{t}{2} \Big\rfloor \Bigg\lceil \frac{\log{k \choose \lfloor t/2\rfloor }}{
- \log\left( 1 - \frac{\lfloor t/2 \rfloor!}{\lfloor t/2 \rfloor^{\lfloor t/2 \rfloor}} \right)}\Bigg\rceil
\cdot r_P\left(\Bigg\lceil \frac{k}{\lfloor t/2 \rfloor} \Bigg\rceil, t-2\right) \; . 
\]
\end{theorem}

In particular, it follows from Theorem~\ref{thrm:redundancy} that for any fixed $t$,
\[
r_B(k,t) = O\left( \sqrt{k} \cdot \log(k) \right) \; . 
\]

\section{Bounds on the Parameters of PIR and Batch Codes with Restricted Size of Reconstruction Sets}
\label{sec:bounds-restricted}

In~\cite{Zhang-Skachek}, the authors study linear batch codes with restricted size of reconstruction sets. 
They aim at refining the Singleton bound for that case by using ideas in~\cite{Yekhanin2012} and subsequent works. 
Note, however, that these ideas cannot be applied directly, because in LRCs there are dependencies between 
different coded symbols, and expurgation of the code in the proof of the bound~(\ref{eq:bound-gopalan})
(and similar bounds) uses these dependencies. Therefore, the authors of~\cite{Zhang-Skachek}
consider a query of $t$ copies of the same symbol (for example, $(x_1, x_1, \cdots, x_1)$), and show that  
the symbols in different reconstruction sets possess certain dependencies. By using this property, they apply 
an expurgation technique similar to that of~\cite{Yekhanin2012, RMV2014, RPDV2014, wang-zhang}, and obtain 
the following relation on the parameters of batch codes with size of reconstruction sets restricted to $r$. 
The proof actually only assumes property $\cA_t$, and therefore it is directly applicable to PIR codes as well.  

\begin{theorem}[\cite{Zhang-Skachek}]
Let $\code$ be a linear $[n,k,t]_q$-batch code (or PIR code) with the 
size of reconstruction sets restricted to $r$. Then, it holds: 
\begin{equation}
n \geq k+d+(t-1)\left(\left\lceil\frac{k}{rt-t+1}\right\rceil-1\right)-1 \; . 
\label{eq:main-statement}
\end{equation}
\label{thrm:batch}
\end{theorem}

Now, observe that if the $[n,k,t]_q$-batch code (or PIR code) allows for reconstruction of any batch of $t$ symbols, 
then it also allows for reconstruction of any batch of $\beta$ symbols, $1 \le \beta \le t$. Therefore, expression~(\ref{eq:main-statement}) in Theorem~\ref{thrm:batch} can be adjusted as follows:   
	
\begin{equation}
\label{bound} 
n\geq k+d+\max_{1\leq \beta\leq t, \beta \in \nn}\left\{(\beta-1)\left(\left\lceil\frac{k}{r\beta-\beta+1}\right\rceil-1\right)\right\} -1\; .
\end{equation}

If the code is systematic, then there is always a reconstruction set of size $1$ for one of the queried symbols. In that case, the last expression can be rewritten as:
\begin{equation}
n\geq k+d+\max_{2\leq \beta\leq t, \beta \in \nn}\Bigg\{(\beta-1)\left(\left\lceil\frac{k}{r\beta-\beta-r+2}\right\rceil-1\right) \Bigg\} -1 \; .
\label{eq:systematic}
\end{equation}
	
The reader can refer to~\cite{Zhang-Skachek} for the full proofs. 

\begin{example}[\cite{Zhang-Skachek}] 
Take $r=2$ and $t=\beta=2$. Then, the bound in~(\ref{eq:systematic}) is attained with equality by the linear systematic codes of minimum distance 2, defined as follows: 
\begin{itemize}
\item $y_i=x_i$ for $1\leq i\leq k$, and $y_j=x_{2(j-k)-1}+x_{2(j-k)}$ for $k+1\leq j\leq k+k/2$, when $k$ is even,
\item $y_i=x_i$ for $1\leq i\leq k$, $y_j=x_{2(j-k)-1}+x_{2(j-k)}$ for $k+1\leq j\leq k+(k-1)/2$, and $y_{k+(k+1)/2}=x_k$, when $k$ is odd.
\end{itemize}
In that case, $d=2$, and we obtain 
\begin{eqnarray*}
& n = k + k/2 \quad & \mbox{ if $k$ is even } \; , \\
& n = k + (k+1)/2 \quad & \mbox{ if $k$ is odd } \; .
\end{eqnarray*}
In both cases, the bound~(\ref{eq:systematic}) is attained with equality for all $k \ge 1$. 
\end{example}

\begin{example}
Consider the code $\code$ in Example~\ref{ex:simplex}, which was studied in \cite{WKC2015}. 
As discussed, $\code$ is a linear $[7,3,4]_2$-batch code, with the size of reconstruction sets at most $r=2$. 
Its minimum Hamming distance is $d=4$. We pick $\beta=2$, and observe that
the right-hand side of equation~(\ref{eq:systematic}) can be re-written as 
\begin{equation*}
 3 + 4 + (2-1)\left(\left\lceil\frac{3}{2\cdot 2- 2 - 2 + 2}\right\rceil-1\right) - 1 = 7 \; ,
\end{equation*}
and therefore the bound in~(\ref{eq:systematic}) is attained with equality for the choice $\beta = 2$. 
The code $\code$ in Example~\ref{ex:simplex} is optimal with respect to that bound. 
We note, however, that general simplex codes (of larger length) do not attain~(\ref{eq:systematic}) 
with equality. 
\end{example}
\medskip 

A slight improvement to the above bounds for both batch and PIR codes can be obtained, if one considers 
simultaneously reconstruction sets for, say, two queried batches $(x_1, x_1, \ldots, x_1)$ and $(x_2, x_2, \ldots, x_2)$,
and studies intersections of their reconstruction sets. The analysis along those lines was done in~\cite{Zhang-Skachek}, and the following result was derived. 

Assume that 
\begin{equation}
k \geq 2(rt-t+1)+1 \; . 
\label{eq:k-range}
\end{equation}
Denote
\begin{eqnarray*}
\sAA & = & \sAA(k, r, d, \beta, \epsilon) 
\; \triangleq \; k + d + (\beta-1)\left(\left\lceil\frac{k+\epsilon}{r\beta-\beta+1}\right\rceil-1 \right)-1 \; , \\
\sBB & = & \sBB(k, r, d, \beta, \lambda) 
\; \triangleq \; k + d + (\beta-1)\left(\left\lceil\frac{k+\lambda}{r\beta-\beta+1}\right\rceil-1\right)-1 \; , \\
\sCC & = & \sCC(k, r, \beta, \lambda, \epsilon) \; \triangleq \; (r\beta-\lambda+1)k-{k\choose 2}(\epsilon-1) \; .
\end{eqnarray*}

\begin{theorem}[\cite{Zhang-Skachek}]
Let $\code$ be a linear $[n,k,t]$-batch code over $\ff$ with the minimum distance $d$ and size of reconstruction sets at most $t$. 
Then,
\begin{equation}
\label{boundn} n\geq 
\max_{\beta\in \nn \cap \left[1,\min\left\{t,\big\lfloor\frac{k-3}{2(r-1)}\big\rfloor\right\}\right]} \left\{ \max_{\epsilon,\lambda\in\nn \cap[1,r\beta-\beta]} \left\{ \min \left\{\sAA,\sBB,\sCC\right\}\right\} \right\} \; . 
\end{equation}
\end{theorem}

\section{Open Questions}
\label{sec:open-questions}

Below, we list some open questions related to batch and PIR codes.  
\begin{enumerate}
\item
Derive tighter bounds on the length or redundancy of batch and PIR codes, in particular, for small alphabet size, for large values of $t$, or for bounded values of $r$ .  
\item
Construct new optimal or sub-optimal batch and PIR codes. 
\item
Do non-linear batch (or PIR) codes have better parameters than their best linear counterparts? 
\item
Do non-systematic batch (or PIR) codes have better parameters than their best systematic counterparts? 
\item
Propose batch and PIR codes that allow for efficient reconstruction algorithms.  
\end{enumerate}


\acknowledgement{
The material in this survey has benefited a lot from discussions of the author with his students and colleagues, including 
Venkatesan Guruswami, Camilla Hollanti, Helger Lipmaa, Sushanta Paudyal, Eldho Thomas, Alexander Vardy, Hui Zhang and 
Jens Zumbr\"agel. This work is supported in part by the grants PUT405 and IUT2-1 from the Estonian Research Council and by the EU COST Action IC1104.
}

\end{document}